\documentclass[twocolumn,tighten]{aastex631}
\usepackage{color}

\usepackage{multirow}
\shortauthors{Sano et al. (2021)}
\shorttitle{Discovery of a wind-blown bubble associated with the SNR G346.6$-$0.2}

\begin{document}
\title{Discovery of a Wind-Blown Bubble Associated with the Supernova Remnant G346.6$-$0.2:\\A Hint for the Origin of Recombining Plasma}

\author[0000-0003-2062-5692]{H. Sano}
\affiliation{Division of Science, National Astronomical Observatory of Japan, Mitaka, Tokyo 181-8588, Japan; hidetoshi.sano@nao.ac.jp}

\author[0000-0002-8152-6172]{H. Suzuki}
\affiliation{Department of Physics, Konan University, 8-9-1 Okamoto, Higashinada, Kobe, Hyogo 658-8501, Japan}

\author[0000-0002-0726-7862]{K. K. Nobukawa}
\affiliation{Faculty of Science and Engineering, Kindai University, 3-4-1 Kowakae, Higashi-Osaka, 577-8502, Japan}

\author[0000-0002-4990-9288]{M. D. Filipovi{\'c}}
\affiliation{Western Sydney University, Locked Bag 1797, Penrith South DC, NSW 1797, Australia}

\author[0000-0002-8966-9856]{Y. Fukui}
\affiliation{Department of Physics, Nagoya University, Furo-cho, Chikusa-ku, Nagoya 464-8601, Japan}

\author[0000-0003-1169-1954]{T. J. Moriya}
\affiliation{Division of Science, National Astronomical Observatory of Japan, Mitaka, Tokyo 181-8588, Japan; hidetoshi.sano@nao.ac.jp}
\affiliation{School of Physics and Astronomy, Faculty of Science, Monash University, Clayton, Victoria 3800, Australia}

\begin{abstract}
We report on CO and \ion{H}{1} studies of the mixed-morphology supernova remnant (SNR) G346.6$-$0.2. We find a wind-blown bubble along the radio continuum shell with an expansion velocity of $\sim10$~km~s$^{-1}$, which was likely formed by strong stellar winds from the high-mass progenitor of the SNR. The radial velocities of the CO/\ion{H}{1} bubbles at $V_\mathrm{LSR} = -82$--$-59$~km~s$^{-1}$ are also consistent with those of shock-excited 1720~MHz OH masers. The molecular cloud in the northeastern shell shows a high-kinetic temperature of $\sim60$~K, suggesting that shock-heating occurred. The \ion{H}{1} absorption studies imply that G346.6$-$0.2 is located on the far side of the Galactic center from us, and the kinematic distance of the SNR is derived to be $11.1_{-0.3}^{+0.5}$~kpc. We find that the CO line intensity has no specific correlation with the electron temperature of recombining plasma, implying that the recombining plasma in G346.6$-$0.2 was likely produced by adiabatic cooling. With our estimates of the interstellar proton density 280 cm$^{-3}$ and gamma-ray luminosity $< 5.8 \times 10^{34}$ erg s$^{-1}$, the total energy of accelerated cosmic rays $W_{\rm p} < 9.3 \times 10^{47}$ erg is obtained. A comparison of the age--$W_{\rm p}$ relation to other SNRs suggests that most of the accelerated cosmic rays in G346.6$-$0.2 have been escaped from the SNR shell.
\end{abstract}

\section{Introduction}
Over the last three decades, X-ray spectroscopic observations using CCD detectors allowed us to reveal detailed properties and evolution of shocked thermal plasma in supernova remnants (SNRs). Because the heavy elements in both the supernova ejecta and the interstellar medium (ISM) are initially in low-ionization or neutral before the shock-heating, it is widely accepted that plasma in young SNRs can be explained as nonequilibrium ionization (NEI, a.k.a. ionizing plasma). After a long time of evolution, plasma in middle-aged SNRs reaches collisional ionization equilibrium (CIE). This standard evolutional scenario, however, has been questioned since the first discovery of recombining (overionized) plasma where heavy nuclei are stripped of more electrons than they would be if in the CIE state \citep{2002ApJ...572..897K}. Although the recombining plasma has been detected from more than a dozen SNRs \citep[see a review of][]{2020AN....341..150Y}, the origin of the peculiar plasma is still under debate.

Because the recombining plasma is characterized by lower electron temperature than the ionization temperature, rapid electron cooling or increasing ionization state of nuclei is needed to produce the plasma state. Two primary scenarios are describing the origin of the recombining plasma in SNRs: adiabatic cooling and thermal conduction scenarios \citep[cf.][]{2020AN....341..150Y}. For the adiabatic cooling scenario (a.k.a. rarefaction scenario), rapid electron cooling occurs when the supernova shockwaves breakout from a dense circumstellar matter (CSM) into the surrounding low-density region such as a wind-blown bubble \citep[e.g.,][]{1989MNRAS.236..885I,1994ApJ...437..770M}. 
For the thermal conduction scenario, on the other hand, rapid electron cooling is caused by shock-interactions with cold/dense molecular clouds via thermal conduction \citep[e.g.,][]{2002ApJ...572..897K,2005ApJ...631..935K}. In fact, the lower electron temperature of recombining plasma is found toward the shocked molecular clouds in the SNRs IC443, W28, W44, and W49B \citep[][]{2017ApJ...851...73M,2018PASJ...70...35O,2020ApJ...890...62O,2021arXiv210612009S}. To distinguish the two primary scenarios, therefore, it is essential to investigate the spatial distribution of plasma conditions and their relation to the ISM.

G346.6$-$0.2 is one of the mixed-morphology SNRs that are classified as radio shell morphology with centrally filled thermal X-rays \citep{1998ApJ...503L.167R}. The small apparent diameter of radio shell $\sim8'$ is consistent with a large distance from us \citep[$\sim5.5$--11 kpc,][]{1998AJ....116.1323K,1993AJ....105.2251D} and its old age \citep[$\sim12$--16~kyr,][]{2011MNRAS.415..301S,2013PASJ...65....6Y,2017ApJ...847..121A}. The thermal X-rays are well reproduced by the recombining plasma model \citep{2013PASJ...65....6Y,2017ApJ...847..121A}. According to \cite{2017ApJ...847..121A}, the recombining plasma likely arises from adiabatic cooling by comparing time scales of thermal conduction and adiabatic cooling. The authors also found additional hard X-rays comprising either a power-law component with a photon index of $\sim2$ or a thermal component with a temperature of $\sim2.0$~keV. In either case, spatially resolved X-ray spectroscopy presented no significant differences in the absorbing column densities, electron temperature of recombining plasma, and the photon indexes toward the SNR \citep{2017ApJ...847..121A}. 

G346.6$-$0.2 is also believed to be interacting with dense molecular clouds. \cite{1998AJ....116.1323K} discovered several shock-excited 1720~MHz OH masers along with the southern shell of the SNR at a velocity range from $-79.3$ to $-74.0$ km s$^{-1}$, implying that the SNR is physically associated with molecular clouds at the same velocities. Subsequent infrared studies revealed the presence of shocked H$_2$ emission in the southern rim and possibly in the northern shell \citep{2006AJ....131.1479R,2011ApJ...742....7A}. Nevertheless, there are no radio-line studies using CO and \ion{H}{1} that can trace the bulk masses of molecular and atomic clouds. Moreover, hadronic gamma-rays---which are produced by interactions between cosmic-ray protons and ISM protons---have never been detected from the SNR \citep{2012AIPC.1505..265E}.

In the present paper, we report on detailed CO and \ion{H}{1} studies using the radio telescopes and interferometers of NANTEN2, Mopra, Atacama Pathfinder Experiment (APEX), Australia Telescope Compact Array (ATCA), and Parkes to reveal the origin of the recombining plasma from G346.6$-$0.2. Section \ref{sec:datasets} describes datasets of CO, \ion{H}{1}, radio continuum, X-rays, and gamma-rays. Section \ref{sec:results} comprises five subsections: Section \ref{subsec:overview} presents distributions of X-ray and radio continuum. Sections \ref{subsec:co_hi}--\ref{subsec:abs} show CO and \ion{H}{1} distributions and their physical properties. Section \ref{subsec:gamma_upper} provides results on the gamma-ray analysis. Discussion and conclusions are given in Sections \ref{sec:discussion} and \ref{sec:conclusions}, respectively.

\begin{deluxetable*}{lccccccc}
\tablewidth{\linewidth}
\tablecaption{Summary of {\it{XMM-Newton}} archive data in G346.6$-$0.2}
\tablehead{
&&&&&\multicolumn{3}{c}{Exposure}\\
\cline{6-8}
\colhead{Obs. ID} & \colhead{$\alpha_{\mathrm{J2000}}$} & \colhead{$\delta_{\mathrm{J2000}}$}  & \colhead{Start Date} & \colhead{End Date} & \colhead{MOS1} & \colhead{MOS2} & \colhead{pn} \\
& \colhead{($^\mathrm{h}$ $^\mathrm{m}$ $^\mathrm{s}$)} & \colhead{($\arcdeg$ $\arcmin$ $\arcsec$)} & \colhead{(yyyy-mm-dd hh:mm:ss)} & \colhead{(yyyy-mm-dd hh:mm:ss)} & \colhead{(ks)} & \colhead{(ks)} & \colhead{(ks)}}
\startdata
0654140101 & 17 10 17.00 & $-40$ 10 22.1 & 2011-03-15 11:51:02 & 2011-03-15 20:13:02 & 29.1 & 28.4 & 27.2 \\
0782080101 & 17 10 09.70 & $-40$ 11 32.5 & 2016-09-04 16:07:23 & 2016-09-05 23:14:02 & 97.5 & 95.2 & 92.6 \\
\enddata
\tablecomments{All exposure times represent the flare-filtered exposure.}
\label{tab1}
\end{deluxetable*}

\section{Datasets}\label{sec:datasets}
\subsection{CO, \ion{H}{1}, and Radio Continuum}\label{subsec:radio}
Observations of $^{12}$CO($J$~=~2--1) line emission at 230~GHz were conducted from August to November in 2008 using the NANTEN2 4-m millimeter/sub-millimeter radio telescope belonging to Nagoya University, which has been installed at Pampa La Bola (4865-m above sea level) in northern Chile. We carried out a Nyquist sampled on-the-ﬂy mapping which covered an area of 2.75 degree$^2$ containing both the SNRs RX~J1713.7$-$3946 and G346.6$-$0.2. Most of the datasets have been published in several papers for RX~J1713.7$-$3946 \citep{2008AIPC.1085..104F,2012ApJ...746...82F,2010ApJ...724...59S,2013ApJ...778...59S,2012MNRAS.422.2230M,2013PASA...30...55M}. The frontend was a 4~K cooled Nb superconductor-insulator-superconductor mixer receiver. The typical system temperature including atmosphere was $\sim250$ K in the single sideband. The backend was an acoustic optical spectrometer with 2048~channels, providing a velocity coverage of 390 km~s$^{-1}$ with a velocity resolution of 0.38 km~s$^{-1}$ at 230~GHz. After convolution using a two-dimensional Gaussian function of $45''$, the final beam size was $\sim 90''$ in the Full-Width-Half-Maximum (FWHM). The absolute intensity was calibrated by observing Orion-KL [($\alpha_\mathrm{J2000}$, $\delta_\mathrm{J2000}$) $=$ ($05^\mathrm{h}35^\mathrm{m}14\fs52$, $-05\arcdeg22\arcmin28\farcs2$)] \citep{1998A&A...335.1049S}. The pointing accuracy was achieved to be better than $\sim 15''$ through two-hourly observations of Jupiter. The typical noise fluctuations toward G346.6$-$0.2 are $\sim0.12$~K at a velocity resolution of 1 km~s$^{-1}$.

We also used archival datasets of $^{12}$CO($J$~=~1--0) and $^{13}$CO($J$~=~1--0, 2--1) line emission to derive the physical properties of molecular clouds: e.g., mass, kinetic temperature, and number density of molecular hydrogen. The $^{12}$CO($J$~=~1--0) and $^{13}$CO($J$~=~1--0) data are from the Mopra Southern Galactic Plane CO Survey Data Release 3 \citep[DR3,][]{2018PASA...35...29B} using the Mopra 22-m radio telescope, and the $^{13}$CO($J$~=~2--1) data are from the Structure, Excitation and Dynamics of the Inner Galactic Interstellar Medium survey \citep[SEDIGISM;][]{2021MNRAS.500.3064S} using the APEX 12-m radio telescope. The angular resolution is $\sim 36''$ for the $^{12}$CO($J$~=~1--0) and $^{13}$CO($J$~=~1--0) data; and $\sim 30''$ for the $^{13}$CO($J$~=~2--1) data. To compare the $^{12}$CO($J$~=~2--1) data obtained with NANTEN2, we smoothed all the datasets to match the FWHM of $90''$ using a two-dimensional Gaussian function. The typical noise fluctuations are $\sim 0.18$~K for the $^{12}$CO($J$~=~1--0) and $^{13}$CO($J$~=~1--0) data; and $\sim 0.06$~K for the $^{13}$CO($J$~=~2--1) data at a velocity resolution of 1 km~s$^{-1}$ for each.

The \ion{H}{1} line data at 1.4~GHz were provided by the Southern Galactic Plane Survey \citep[SGPS;][]{2005ApJS..158..178M} using ATCA combined with the Parkes 64-m radio telescope. The combined beam size is $130'' \times 130''$ and a velocity resolution of 0.82~km~s$^{-1}$. The typical noise fluctuations are $\sim 1.3$ K at a velocity resolution of 1~km~s$^{-1}$.

The radio continuum data at 843~MHz are from the Molonglo Observatory Synthesis Telescope (MOST) supernova remnant catalogue \citep[MSC,][]{1996A&AS..118..329W}. The angular resolution is $\sim 43''$ and the typical noise fluctuations are $\sim 2$ mJy beam$^{-1}$.

\subsection{X-rays}\label{subsec:xrays}
We used archival datasets obtained with {\it{XMM-Newton}} to present a map of X-ray recombining plasma in G346.6$-$0.2, for which the observation IDs are 0654140101 (PI: C.-Y. Ng) and 0782080101 (PI: K. Auchettl). We used the XMM-Newton Science Analysis System \citep[SAS,][]{2004ASPC..314..759G} version 19.1.0 and HEAsoft version 6.28 to analyze both the EPIC-MOS and EPIC-pn datasets which were obtained using the full-frame mode with the thick filter. Table \ref{tab1} lists the details of the observations. We reprocessed the Observation Data Files (ODF) for each pointing, following a standard procedure provided as the XMM-Newton Extended Source Analysis Software \citep[ESAS,][]{2008A&A...478..575K}. The effective exposures after filtering soft proton flares are also shown in Table \ref{tab1}. The total exposure time is $\sim 370$ ks. To subtract the Quiescent Particle Background (QPB) from the X-ray map, we run ``mos-$/$pn-back'' and ``mos-$/$pn-filter'' procedures. We also used the procedure ``merge\_comp\_xmm'' to combine the two-pointing data. Note that we excluded some CCD chips or quadrants which are affected by strong stray light \citep[see also Figure 2 of][]{2017ApJ...847..121A}. We then applied an adaptive smoothing by using the ``adopt\_merge'' procedure, where the pixel size and smoothing counts were $6''$ and 300 counts, respectively. Finally, we produced a QPB-subtracted, exposure-corrected, and adaptively smoothed image in the energy band of 0.5--7.0 keV.

\subsection{Gamma-Rays}\label{subsec:gammarays}
In order to get information of particle acceleration, we also studied the gamma-ray emission around this source with {\it Fermi}-LAT. We extracted all the available data for the circle region with a radius of $20^\circ$ centered at the source position from the Pass 8 database\footnote{\url{https://fermi.gsfc.nasa.gov/cgi-bin/ssc/LAT/LATDataQuery.cgi}}. The energy range for the data extraction is 100~MeV to 300~GeV.

The tool and databases used in the data reduction/analysis are {\it Fermitools} (v1.2.23)\footnote{\url{https://github.com/fermi-lat/Fermitools-conda/}}, the Instrumental Response File version P8R3\_SOURCE\_V2, the {\it Fermi} source list gll\_psc\_v22.fit (4FGL catalog), the Galactic diffuse background model gll\_iem\_v07.fits, and the isotropic background model (instrumental and extragalactic) iso\_P8R3\_SOUCE\_V2.txt.

The standard event selection is applied to the retrieved data set: from the SOURCE class, both FRONT and BACK section events are extracted ({\it evclass=128 evtype=3}). Events with zenith angles larger than 90$^\circ$ are rejected from the analysis in order to prevent contamination from the Earth's bright limb. The energy dispersion correction is enabled for all the model components but the isotropic background model\footnote{\url{https://fermi.gsfc.nasa.gov/ssc/data/analysis/documentation/Pass8_edisp_usage.html}}.

\begin{figure*}[]
\begin{center}
\includegraphics[width=\linewidth,clip]{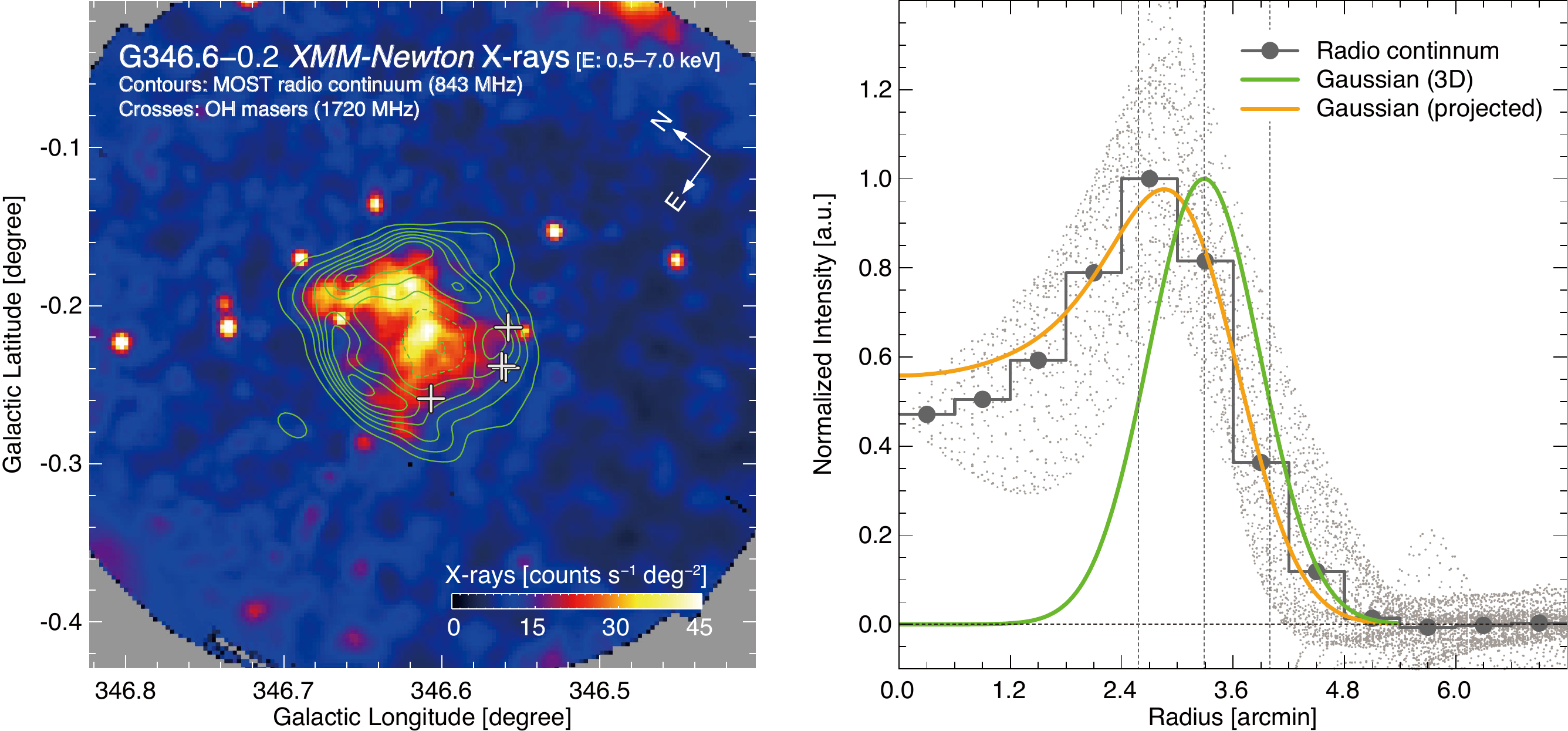}
\caption{{\it{Left panel}}: Map of the {\it{XMM-Newton}} X-ray flux toward the SNR G346.6$-$0.2 ($E$: 0.5--7.0~keV). The superposed contours indicate the 843~MHz radio continuum from the MOST Supernova Remnant Catalogue \citep[][]{1996A&AS..118..329W}. The lowest contour level and the contour intervals are 20 and 40 mJy beam$^{-1}$, respectively. The crosses represent the positions of 1720~MHz OH masers \citep{1998AJ....116.1323K}. {\it{Right panel}}: Radial profiles of the radio continuum, centered at ($l$, $b$) = ($346\fdg62$, $-0\fdg22$). The small dots represent all the data points for the radio continuum. The black steps with large filled circles indicate averaged values of radio continuum at each annulus. The green line represents the three-dimensional Gaussian distribution and the orange line represents its projected distribution derived by the least-squares fitting (see the text). The vertical dashed lines indicate best-fit values of radius and ranges of shell thickness.}
\label{fig1}
\end{center}
\end{figure*}%

\section{Results}\label{sec:results}
\subsection{Overview of X-rays and Radio Continuum}\label{subsec:overview}
Figure \ref{fig1} left panel shows the overlay map of X-rays and radio continuum toward G346.6$-$0.2. As presented in previous studies, the diffuse X-ray emission produced by recombining plasma shows the centrally filled distribution inside the radio continuum shell \citep{2013PASJ...65....6Y,2017ApJ...847..121A}. The highest value of X-ray flux found in the center and northwestern regions, while the northeast, west, and southeast regions inside the shell are dim in the X-rays. Note that this trend is not caused by interstellar absorption because the spatially resolved X-ray spectroscopy indicated almost uniform absorbing column densities within the shell \citep{2017ApJ...847..121A}. We also find that the positions of the shock-excited OH masers spatially correspond to the outer boundary of the diffuse X-ray emission. There are a dozen of X-ray point sources, almost all of which have been analyzed by \cite{2017ApJ...847..121A} and will not be discussed further in this paper.

To derive the apparent diameter and shell thickness of the radio shell quantitatively, we fitted its radial profile using a three-dimensional spherical shell with a Gaussian function $F(r)$:
\begin{eqnarray}
F(r) =  A \exp[-(r - r_0\bigr)^2/2\sigma ^2]
\label{eq1}
\end{eqnarray}
where $A$ is a normalization of the Gaussian function, $r_0$ is the shell radius in units of arcmin, and $\sigma$ is the standard deviation of the Gaussian function in units of arcmin. We first fitted the radial profile by changing an original position around the geometric center of the SNR. This process gave the central position of the shell to be ($l$, $b$) = (346\fdg62, $-$0\fdg22) with the minimum chi-square value of the least-squares fitting. Figure \ref{fig1} right panel shows the radial profiles of the radio continuum and X-rays. We derived the shell radius $r_0$ of $3\farcm29 \pm 0\farcm05$ and thickness of $1\farcm41 \pm 0\farcm11$ as the best-fit parameters, where the shell thickness is defined as the FWHM of the Gaussian function or $2\sigma \sqrt{2 \ln 2}$.

\begin{figure*}[]
\begin{center}
\includegraphics[width=\linewidth,clip]{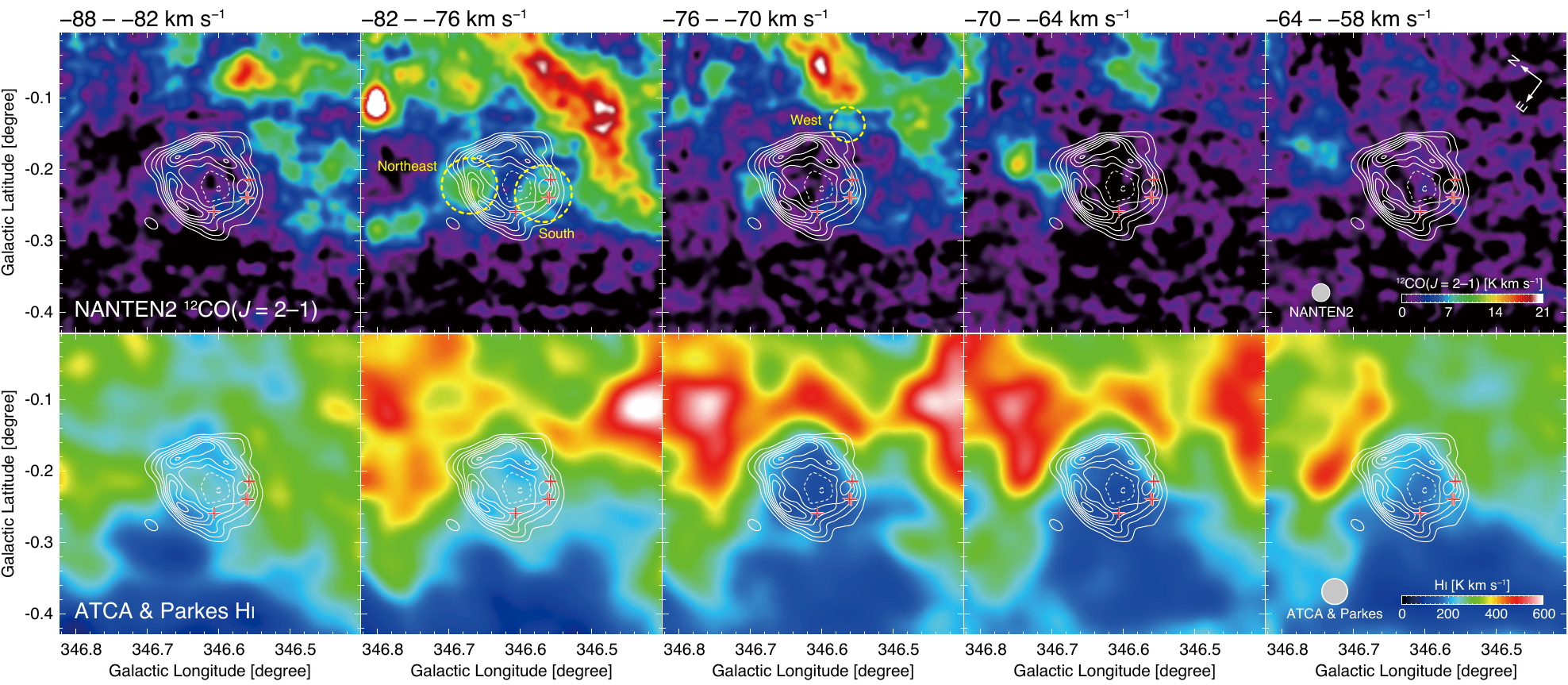}
\caption{Velocity channel distributions of the NANTEN2 $^{12}$CO($J$~=~2--1) ({\it{top panels}}) and ATCA \& Parkes \ion{H}{1} ({\it{bottom panels}}). The superposed contours and crosses are the same as shown in Figure~\ref{fig1}. Each panel shows CO or \ion{H}{1} distributions every 6 km s$^{-1}$ in a velocity range from $-88$ to $-58$ km s$^{-1}$. The CO clouds in the northeast, south, and possibly west are also enclosed by dashed yellow circles.}
\label{fig2}
\end{center}
\end{figure*}%

\begin{figure*}[]
\begin{center}
\includegraphics[width=\linewidth,clip]{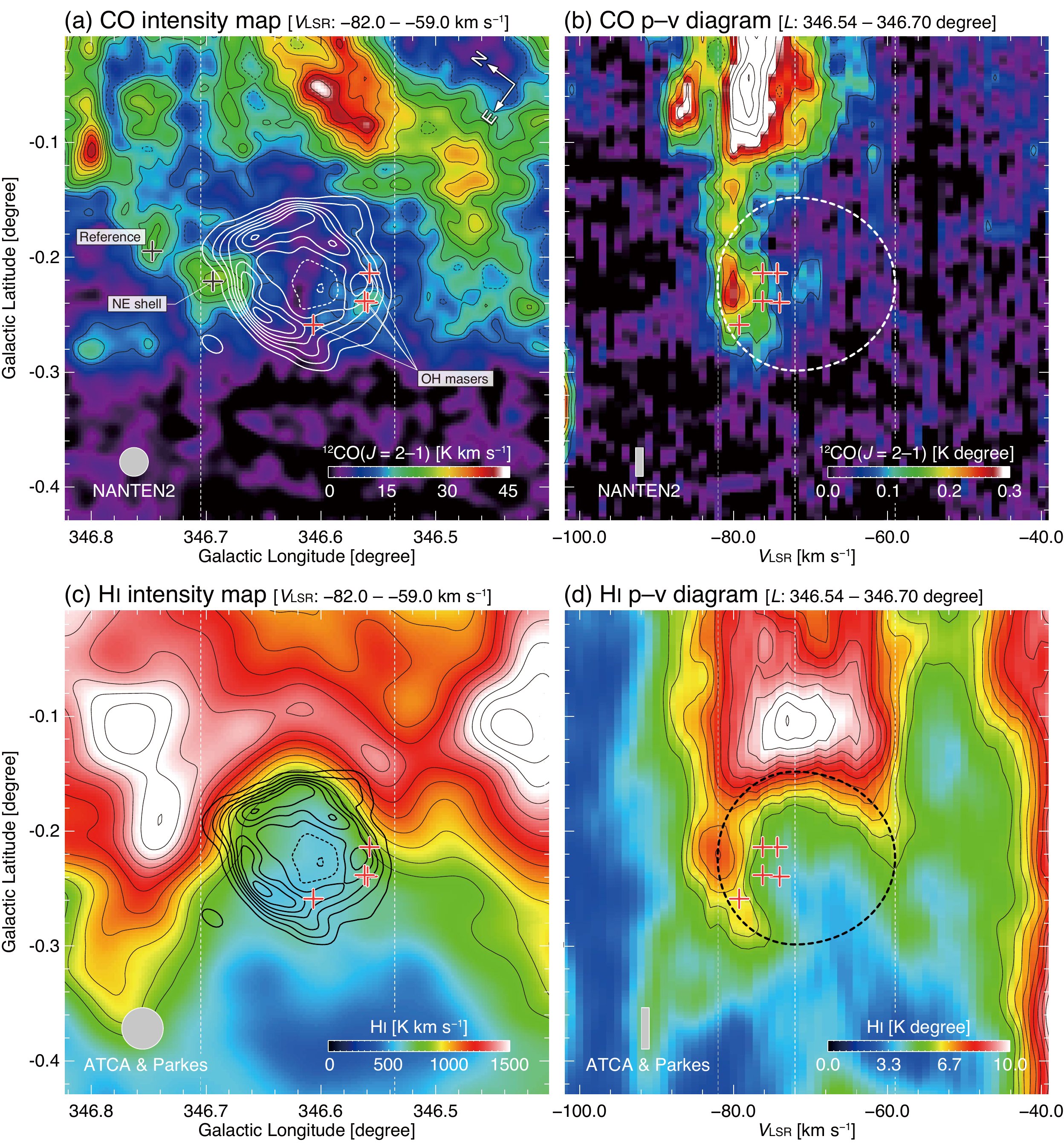}
\caption{Integrated intensity maps and position--velocity (p--v) diagrams of $^{12}$CO($J$~=~2--1) ({\it{top panels}}) and \ion{H}{1} ({\it{bottom panels}}). The integration range is from $-82$ to $-59$ km s$^{-1}$ in the velocity for each intensity map; and from 346\fdg54 to 346\fdg70 in the galactic longitude for each p--v diagram. The solid contours represent the radio continuum whose contour levels are the same as shown in Figure~\ref{fig1}. The lowest contour level and the contour intervals are 12 and 3 K~km$^{-1}$ for the CO intensity map; 800 and 80 K~km$^{-1}$ for the \ion{H}{1} intensity map; 0.08 and 0.04 K degree for the CO p--v diagram; 5.5 and 0.5 K degree for the \ion{H}{1} p--v diagram. The red crosses represent the positions of 1720~MHz OH masers \citep{1998AJ....116.1323K}. The CO peaks ``NE-shell'' and ``Reference'' discussed in Section \ref{subsec:lvg} are also indicated in (a). The dashed circles in the p--v diagrams indicate the boundaries of the CO and \ion{H}{1} cavities (see the text).}
\label{fig3}
\end{center}
\end{figure*}%

\subsection{Distributions of CO and \ion{H}{1}}\label{subsec:co_hi}
Figure \ref{fig2} shows the velocity channel maps of CO and \ion{H}{1} toward G346.6$-$0.2. In the present paper, we focus on only the velocity range from $-88$ to $-58$ km s$^{-1}$, which is roughly consistent with the radial velocities of the shock-excited OH masers \citep{1998AJ....116.1323K}. We find several molecular clouds overlapping with the radio shell, especially toward the northeast, south, and possibly in the west as shown by dashed yellow circles. The south CO clouds show a good spatial correspondence with the positions of the shock-excited OH masers. In the \ion{H}{1} channel maps, we find a cavity-like distribution toward the radio shell in the velocity range from $-82$ to $-58$ km s$^{-1}$. The rich \ion{H}{1} clouds are located along with the northwestern half of the SNR, while the southeastern half shows a blowout structure.

Figures \ref{fig3}a and \ref{fig3}c show the integrated intensity maps of CO and \ion{H}{1} at $V_\mathrm{LSR} = -82$--$-58$ km s$^{-1}$. The spatial relation between the radio shell and CO/\ion{H}{1} clouds is clearer than the velocity channel maps. We note that the CO intensity of the south clouds overlapping with the shock-excited OH masers are weaker than that of the molecular cloud laying on the edge of the northeastern shell (refer to as ``NE-shell cloud''). Moreover, there are no dense CO/\ion{H}{1} clouds toward the center of the SNR where the X-ray recombining plasma is strongly detected.

Figures \ref{fig3}b and \ref{fig3}d show the position--velocity (p--v) diagrams of CO and \ion{H}{1}. We find a cavity-like structure in the p--v diagram of \ion{H}{1}, whose velocity range is from $-82$ to $-58$ km s$^{-1}$. Note that all the shock-excited OH masers are located inside the velocity range of the cavity. It is noteworthy that the spatial extent of the cavity in the Galactic longitude direction is roughly consistent with the apparent diameter of the radio shell. Moreover, the shock-excited OH masers are located on the low-density regions (or surface) of the molecular clouds in the velocity space (see Figure \ref{fig3}b). 

\subsection{LVG Analysis}\label{subsec:lvg}
To estimate the kinetic temperature and density of molecular clouds, we performed the Large Velocity Gradient analysis \citep[LVG, e.g.,][]{1974ApJ...189..441G,1974ApJ...187L..67S}. The LVG analysis calculates the radiative transfer of multi-transitions of molecular line emission, assuming a spherically isotropic cloud with uniform kinetic temperature, photon escape probability, and a velocity gradient of $dv/dr$. Here, $dv$ is the Half-Width-Half-Maximum (HWHM) of CO line profiles obtained by the least-square fitting using a single Gaussian function, and $dr$ is a cloud radius. To test the shock-heating and/or compression, we selected two molecular clouds which are significantly detected both in the $^{12}$CO($J$~=~2--1) and $^{13}$CO($J$~=~1--0, 2--1) line emission as well as free from extra heating sources such as infrared and/or young stellar objects\footnote{Although the southern radio shell is overlapping with both the OH masers and weak CO emission, there are no molecular clouds that are detected in all molecular lines as well as free from any extra heating sources.}. One of the clouds is the NE-shell cloud which is a candidate for the post-shocked cloud. The other was selected as a reference which is located outside of the radio shell (refer to as ``reference cloud''). Therefore, it is expected that the shock-heating and/or compression will be seen only in the NE-shell cloud if the cloud is interacting with the SNR.

\begin{figure}[]
\begin{center}
\includegraphics[width=\linewidth,clip]{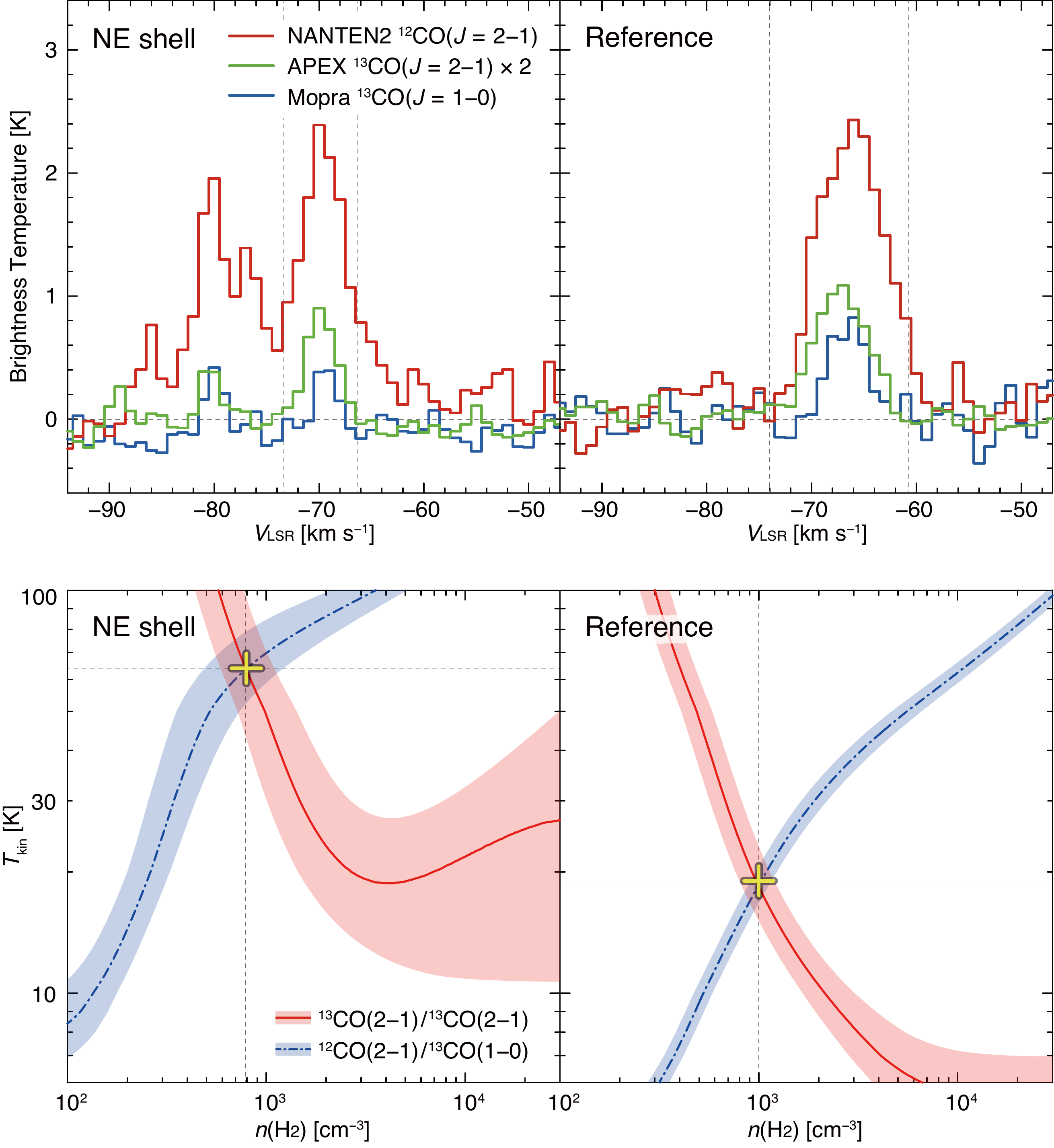}
\caption{{\it{Top panels}}: CO intensity profiles toward the CO peaks of the NE-shell and Reference clouds. Each CO data was smoothed to match the beam size $90''$ and the velocity resolution of 0.5 km s$^{-1}$. {\it{Bottom panels}}: Results of LVG analysis on the number density of molecular hydrogen $n$(H$_2$) and the kinetic temperature $T_{\mathrm{kin}}$ for each CO peak. The red lines and blue dash-dotted lines indicate the intensity ratios of $^{13}$CO($J$~=~2--1) / $^{13}$CO($J$~=~2--1) and $^{12}$CO($J$~=~2--1) / $^{13}$CO($J$~=~1--0), respectively. The shaded areas in red and blue indicate $1\sigma$ error regions for each intensity ratio. Yellow crosses represent the best-fit values of $n$(H$_2$) and $T_{\mathrm{kin}}$ for each CO peak.}
\label{fig4}
\end{center}
\end{figure}%

\begin{figure*}[]
\begin{center}
\includegraphics[width=\linewidth,clip]{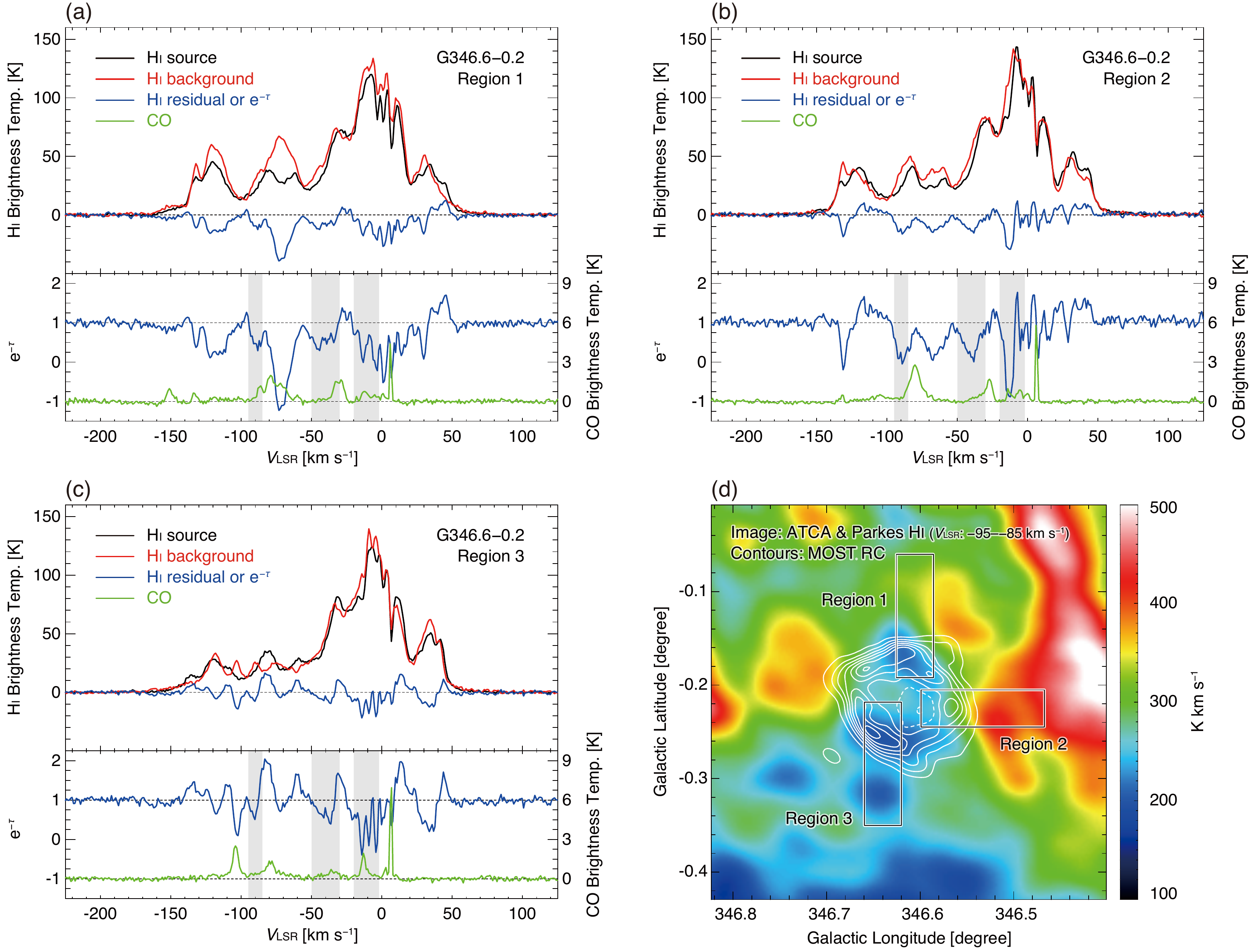}
\caption{(a--c) \ion{H}{1} and CO spectra of three rectangular regions 1--3 toward G346.6$-$0.2. Top half of each panel shows \ion{H}{1} profiles of source region (black), background region (red), and the residual (blue). Bottom half of each panel indicates \ion{H}{1} absorption profiles of $e^{-\tau}$ (blue) and $^{12}$CO($J$~=~1--0) line emission (green). The gray shaded areas represent velocity ranges that show \ion{H}{1} absorption features in all three regions. (d) Integrated intensity of \ion{H}{1} superposed on the radio continuum contours. The integrated velocity range is from $-95$ to $-85$ km~s$^{-1}$. The contour levels of radio continuum are the same as shown in Figure~\ref{fig1}. The three rectangular regions 1--3 are also indicated.}
\label{fig5}
\end{center}
\end{figure*}%

Figure \ref{fig4} upper panels show the CO spectra toward the two clouds.  We note that the $^{12}$CO($J$~=~2--1) profile of the NE-shell cloud shows multiple peaks at $\sim -86$, $-80$, and $-70$ km s$^{-1}$ with a broad component at the velocity range approximately from $-90$ to $-60$ km s$^{-1}$. We estimated $dv/dr = 0.5$ km s$^{-1}$ pc$^{-1}$ for the NE-shell cloud at the $\sim70$ km s$^{-1}$ component, and 1 km s$^{-1}$ pc$^{-1}$ for the reference cloud, assuming the largest expected distance of 11 kpc. In the case of the near-side distance at 5.5 kpc, the values of $dv/dr$ will decrease the only factor of two and will not affect the results significantly. We also used the abundance ratio of [$^{12}$CO/H$_2$] = $5 \times 10^{-5}$ \citep{1987ApJ...315..621B} and the abundance ratio of isotopes $[^{12}\mathrm{CO}/^{13}\mathrm{CO}] = 33$ \citep{1990ApJ...357..477L}. 

Figure \ref{fig4} bottom panels show the results of the LVG analysis toward the NE-shell and reference clouds, which describe a relation between the kinetic temperature $T_\mathrm{kin}$ and the number density of molecular hydrogen $n(\mathrm{H}_2)$. We derived the best-fit parameters of $T_\mathrm{kin} = 64^{+18}_{-14}$ K and $n(\mathrm{H}_2) = 790^{+100}_{-70}$ cm$^{-3}$ for the NE-shell cloud, and $T_\mathrm{kin} = 19^{+3}_{-3}$ K and $n(\mathrm{H}_2) = 1000^{+70}_{-50}$ cm$^{-3}$ for the reference cloud. The values of $n(\mathrm{H}_2)$ are roughly consistent with each other ($\sim1000$ cm$^{-3}$), while the kinetic temperature of the NE-shell cloud is more than three times as high as that of the reference cloud.

\subsection{\ion{H}{1} Absorption Studies}\label{subsec:abs}
To investigate the distance of G346.6$-$0.2, we carried out \ion{H}{1} absorption studies. Following the latest method described by \cite{2010ASPC..438..365L} and \cite{2017ApJ...843..119R}, we selected three regions 1, 2, and 3 toward the SNR (Figure \ref{fig5}d). Figures \ref{fig5}a--\ref{fig5}c show the averaged \ion{H}{1} source, background, and residual (or $e^{-\tau}$) profiles as well as the CO spectra toward the regions 1, 2, and 3, respectively. We find significant absorption features of all \ion{H}{1} spectra at three velocity ranges of $V_\mathrm{LSR} = -95$--$-85$, $-50$--$-30$, and $-20$--$-2$ km s$^{-1}$. The three velocity ranges clearly show negative residuals (or $e^{-\tau} < 1$) without strong CO line emission, suggesting that the absorption profiles are not caused by the self-absorption effect due to cold/dense clouds. Moreover, the \ion{H}{1} integrated intensity maps of three velocity ranges show intensity dips of \ion{H}{1} toward the radio shell. Figure \ref{fig5}d shows the \ion{H}{1} integrated intensity map at the velocity range from $-95$ to $-85$ km s$^{-1}$. The \ion{H}{1} intensity dips show a nice spatial correspondence with the radio bright shell. This means that the \ion{H}{1} absorption features due to the radio continuum emission from the SNR found at least the velocity range from $-95$ to $-2$ km s$^{-1}$. The interpretation of this observational result will be discussed in Section \ref{subsec:distance} in detail.

\subsection{Upper Limit of GeV Gamma-Ray Flux}\label{subsec:gamma_upper}
Here, gamma-ray emission at GeV energies from the source was examined.
The analysis region was a 20$^\circ$-radius circle centered at the SNR position.
In order to get the energy flux of the source, a binned maximum likelihood analysis on the spatial and energy distributions of the data was conducted.
The spatial bin size was $0^\circ_\cdot2 \times 0^\circ_\cdot2$. The data were divided into 35 logarithmic energy bins in the energy range of 100~MeV--300~GeV.
All the sources in the 4FGL catalog within our analysis region were considered in the likelihood analysis, and the spectral parameters of only those within a $8^\circ_\cdot5$-radius circle were treated as free parameters.
For the sources included in the 4FGL catalog, the spectral models in the catalog were used.
For G346.6$-$0.2 itself, a simple power-law model was assumed.

The maximum likelihood analysis was conducted to search for the model configuration which best represents the data. This procedure was repeated with decreasing fit tolerance values until the difference between the resultant likelihood and that in the last trial becomes less than unity. Then, we obtained the best-fit source flux. The test-statistic (TS) for this source was, however, found to be 12.4, which corresponds to a $\approx 3.5 \sigma$ detection. Thus, we concluded that the source detection was marginal or insignificant, and derived only the upper limit of the energy flux as described below.

In the calculation of the upper limit of the energy flux, we considered systematic uncertainties associated with following components: {\it Fermi}-LAT's effective area ($\pm 3\%$) and point spread function ($\pm 5\%$)\footnote{\url{https://fermi.gsfc.nasa.gov/ssc/data/analysis/LAT_caveats.html}}, and the Galactic diffuse background model ($\pm 6\%$; e.g., \citealt{2009ApJ...706L...1A, 2010ApJ...717..372C, 2011ApJ...740L..51T}).
We ran the maximum likelihood analysis several times considering these potential uncertainties, and obtained the energy fluxes and errors of the source in individual runs. The largest value among these was used for estimation of the upper limit. The resultant 95\% upper limit of the energy flux is $4.0 \times 10^{-12}$ erg s$^{-1}$ cm$^{-2}$ in the 1--100~GeV energy range, which is converted to $1.4 \times 10^{34}$ erg s$^{-1}$ at 5.5~kpc or $5.8 \times 10^{34}$ erg s$^{-1}$ at 11~kpc.

\section{Discussion}\label{sec:discussion}
\subsection{Molecular and Atomic Clouds Associated with the SNR G346.6$-$0.2}\label{subsec:associated}
The previous ISM studies of G346.6$-$0.2 suggested the presence of shocked molecular clouds by observing the shock-excited 1720 MHz OH masers at $\sim -70$ km s$^{-1}$ and the near-infrared shocked H$_2$ emission \citep[][]{1998AJ....116.1323K,2006AJ....131.1479R,2011ApJ...742....7A}. However, the shock--interacting neutral gaseous medium---molecular and atomic hydrogen clouds---has not been reported due to a lack of comprehensive studies using CO and \ion{H}{1} emission lines. In the present section, we argue that both the molecular and atomic clouds at $V_\mathrm{LSR} = -82$--$-59$ km s$^{-1}$ are physically associated with the SNR G346.6$-$0.2.

First, we argue that the cavity-like structure in the p--v diagram of \ion{H}{1} (partially CO as well) provides strong support for the shock--interactions with the molecular and atomic clouds at $V_\mathrm{LSR} = -82$--$-59$ km s$^{-1}$. Because such cavity-like structure toward an SNR corresponds to an expanding gas motion, also called the ``wind-blown bubble'', which is thought to be formed by strong winds from the progenitor system of the SNR: e.g., stellar wind from the high-mass progenitor or accretion wind from the progenitor system containing a white dwarf and a companion star \citep[e.g.,][]{1990ApJ...364..178K,1991ApJ...382..204K,1999ApJ...519..314H,1999ApJ...522..487H}. It is noteworthy that the size of such a wind-blown bubble is expected to be consistent with that of the SNR shell because the shock propagation time inside the bubble is very short due to a much-lower density \citep[e.g.,][]{1977ApJ...218..377W}. Moreover, the expansion velocities of wind-blown bubbles were observationally derived to be $\sim3$--13 km s$^{-1}$ in the several Galactic and Magellanic SNRs \citep[e.g.,][]{1989MNRAS.237..277L,2012ApJ...746...82F,2016ApJ...826...34Z,2017JHEAp..15....1S,2018ApJ...867....7S,2019ApJ...881...85S,2018ApJ...864..161K}. In the case of G346.6$-$0.2, the wind-blown bubble shows the expanding velocity of $\sim10$ km s$^{-1}$ and the size of which is roughly consistent with that of the radio continuum shell (see Figure \ref{fig3}). Although the progenitor type of G346.6$-$0.2 is still under debate, the core-collapse origin is thought to be favored \citep[][]{2013PASJ...65....6Y,2017ApJ...847..121A}. We, therefore, propose that the wind-blown bubble at $V_\mathrm{LSR} = -82$--$-58$ km s$^{-1}$ was formed by the strong stellar wind from the high-mass progenitor and is now interacting with supernova shockwaves.

The good correspondences between the CO clouds and shock-excited 1720~MHz OH masers are also consistent with this scenario. In fact, the positions of OH masers spatially coincide with the CO clouds in the southern shell (see Figures \ref{fig2}, \ref{fig3}a, and \ref{fig3}c). The OH masers also lie on the edge of the CO clouds in the velocity space, whose center velocities are in the range of the wind-blown bubble (see Figures \ref{fig3}b and \ref{fig3}d). Because moderate densities of $10^3$--$10^5$ cm$^{-3}$ and temperature of 50--125 K are needed to efficiently emit the shock-excited 1720~MHz OH maser \citep{1976ApJ...203..124E}, the slight velocity offsets between the CO intensity peaks (the densest part of the clouds) and the OH masers are naturally expected (see Figure \ref{fig3}b).

The high-kinetic temperature in the NE-shell cloud provides alternative evidence for the shock-cloud interaction. In Figure \ref{fig4}, we can find that the kinetic temperature of the NE-shell cloud (= post-shocked cloud) is more than three times as high as that of the reference cloud (= pre-shocked cloud). Because there are no extra heating sources except for the supernova shocks, the high temperature of the NE-shell can be understood as the shock-heating. The kinetic temperature of $\sim60$ K in the NE-shell cloud is also consistent with that of shock-heated molecular clouds in the similar mixed-morphology SNRs W44, IC443, and W28 \citep[$T_\mathrm{kin} \sim40$--80 K,][]{1998ApJ...505..286S,1999PASJ...51L...7A,2014A&A...569A..81A}. In addition, the broad velocity component of the NE-shell cloud at $V_\mathrm{LSR} \sim -90$--$-60$ km s$^{-1}$ possibly corresponds to the line-broadening due to the shock acceleration \citep[e.g.,][]{1977ApJ...216..440W,1981ApJ...245..105W,1979ApJ...232L.165D}. Further high spatial resolution and high sensitivity CO observations using the Atacama Large Millimeter/submillimeter Array (ALMA) will confirm the line-broadening in detail. In light of these considerations, we conclude that the CO/\ion{H}{1} clouds as well as the wind-blown bubble in the velocity range from $-82$ to $-59$ km s$^{-1}$ are physically associated with the SNR G346.6$-$0.2.

\subsection{Distance and Age}\label{subsec:distance}
Based on the physical association between the shockwaves and the stellar wind bubble, we here discuss the distance and age of G346.6$-$0.2. We first derived the systemic velocity of the stellar wind bubble as $V_\mathrm{LSR} = -72^{+13}_{-10}$ km s$^{-1}$ from the p--v diagram of \ion{H}{1} (Figure \ref{fig3}d). Here, the systemic velocity is defined as the velocity that shows the largest size of the cavity, and its errors represent to the minimum and maximum velocities of the wind bubble. By adopting the Galactic rotation curve model with the IAU-recommended values of $R_0 = 8.5$ kpc and $\Theta_0 = 220$ km s$^{-1}$ \citep{1986MNRAS.221.1023K,1993A&A...275...67B}, we obtained the kinematic distance of G346.6$-$0.2 as $5.4^{+0.3}_{-0.5}$ kpc for the near side distance and $11.1^{+0.5}_{-0.3}$ kpc for the far side distance.

Next, we argue that the far side distance would be appropriate for the distance to G346.6$-$0.2. In case the SNR is located in the near side distance, we can find the \ion{H}{1} absorption feature only in the velocity range from $-72$ to 0 km s$^{-1}$ because the absorption line can be seen in the foreground gas with respect to the SNR. However, our results show that the absorption features due to the SNR are seen in the velocity range from $-95$ to $-2$ km s$^{-1}$ (see Section \ref{subsec:abs} and Figure \ref{fig5}). We, therefore, conclude that the SNR G346.6$-$0.2 is located on the far side of the Galactic center from us and the appropriate distance is to be $11.1^{+0.5}_{-0.3}$ kpc\footnote{Although the distance is slightly inconsistent with the previous distance of $\sim9$ kpc derived using the $\Sigma$--$D$ relation \citep{1993AJ....105.2251D}, it does not matter considering a large scatter of the relation \citep[e.g.,][]{2014SerAJ.189...25P}.}. We then derived the radius of the SNR to be $10.6^{+0.5}_{-0.3}$ pc.

To estimate the age of G346.6$-$0.2, we calculated the electron density $n_\mathrm{e}$ within the stellar wind bubble. According to \cite{2017ApJ...847..121A}, the electron density with a plasma filling factor $f$ was derived to as $1.4 f^{-0.5} (d/8.3\:\mathrm{kpc})^{-0.5}$ cm$^{-3}$ using the emission measure of the recombining plasma component as well as assuming the electron to ion ratio and the plasma emitting volume (see \citeauthor{2017ApJ...847..121A} \citeyear{2017ApJ...847..121A} for details). We then obtained the electron density $n_\mathrm{e} \sim 1.2$ cm$^{-3}$ at the SNR distance of 11.1 kpc. Because the recombination timescale $n_\mathrm{e} t = (5.3 \pm 0.6) \times 10^{11}$ cm$^{-3}$ s, the elapsed time $t$ after producing the recombining plasma can be estimated to $14 \pm 2$~kyr\footnote{Note that the plasma age of the SNR is expected to have a systematic error within a factor of four in addition to its statistical error of 2 kyr \citep[cf.][]{2021ApJ...914..103S}.}. This values is roughly consistent with the previous age estimation of the SNR \citep[][]{2011MNRAS.415..301S,2013PASJ...65....6Y,2017ApJ...847..121A}. In the present paper, we use $\sim$14~kyr as the age of the SNR G346.6$-$0.2\footnote{According to \cite{1989MNRAS.236..885I}, adiabatic cooling occurs a few hundred years after a supernova explosion. We thus assumed that the plasma age is roughly consistent with the SNR age.}.

\begin{figure*}[]
\begin{center}
\includegraphics[width=\linewidth,clip]{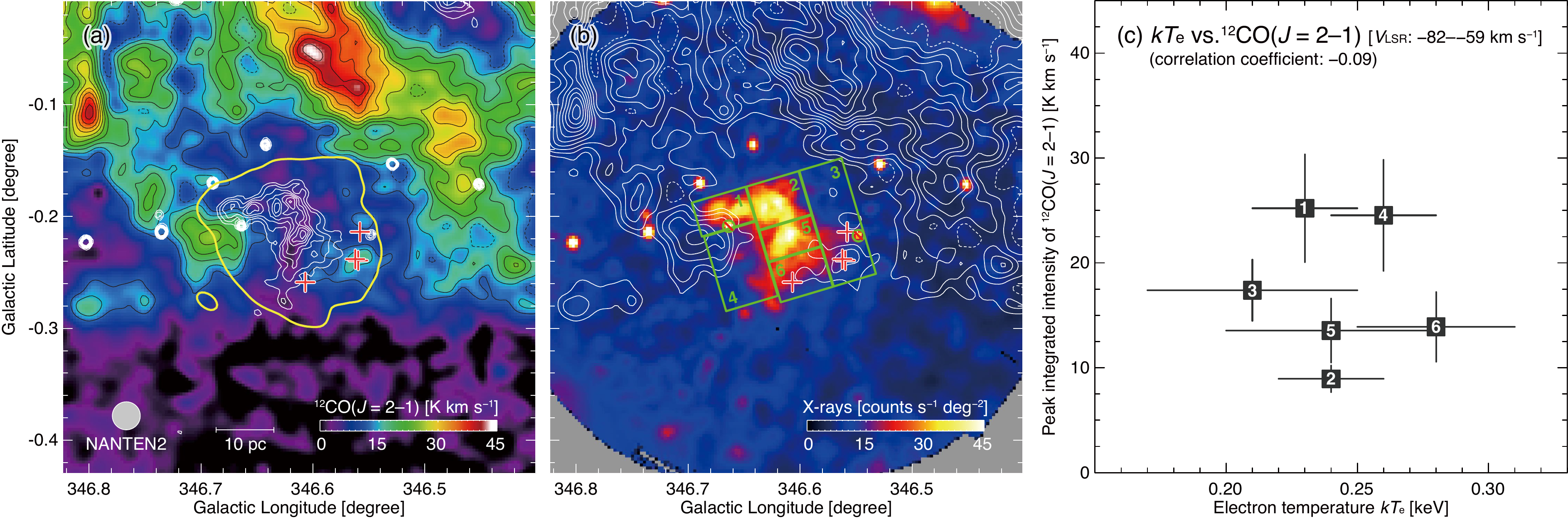}
\caption{(a) Same $^{12}$CO($J$~=~2--1) intensity map as shown in Figure \ref{fig3}(a), but the superposed white contours indicate the {\it{XMM-Newton}} X-ray flux. The lowest contour level and the contour intervals of X-rays are 12 and 3 counts s$^{-1}$ degree$^{-2}$, respectively. The shell boundary of radio continuum is also shown in the yellow contour. The red crosses indicate the positions of 1720~MHz OH masers \citep{1998AJ....116.1323K}. (b) Same as (a) except that the colored image is replaced by the {\it{XMM-Newton}} X-ray flux as shown in Figure \ref{fig1}. The green boxes 1--6 represent the extracted areas of X-ray spectra as defined by \cite{2017ApJ...847..121A}. (c) Scatter plot between the electron temperature $kT_\mathrm{e}$ and the peak integrated intensity of $^{12}$CO($J$~=~2--1) line emission. The vertical error bars represent the standard deviations of CO intensities for each region.}
\label{fig6}
\end{center}
\end{figure*}%

\subsection{Origin of the Recombining Plasma}\label{subsec:rp}
\cite{2017ApJ...847..121A} proposed that the recombining plasma in G346.6$-$0.2 was likely formed by adiabatic cooling because the adiabatic cooling timescale of $\sim12$ kyr is compatible with the age of the SNR. The authors also derived the cooling timescale for thermal conduction to be $\sim500$ kyr, and hence concluded that thermal conduction is most likely not responsible for producing the recombining plasma. In this section, we argue that our ISM results also support the adiabatic cooling scenario as the origin of the recombining plasma in G346.6$-$0.2.

Figures \ref{fig6}a and \ref{fig6}b show the overlay maps of the CO intensity and the X-ray flux. We find a clear spatial anti-correlation between the recombining plasma and shock-interacting molecular clouds especially toward the center of the SNR. Figure \ref{fig6}c shows the scatter plot between the electron temperature $kT_\mathrm{e}$ and the peak integrated intensity of $^{12}$CO($J$~=~2--1) line emission for the regions 1--6 which were defined by \cite{2017ApJ...847..121A}. Although a negative correlation between the CO intensity and $kT_\mathrm{e}$ will be expected for the thermal conduction scenario \citep[][]{2017ApJ...851...73M,2018PASJ...70...35O,2020ApJ...890...62O,2021arXiv210612009S}, we could not find such a correlation (correlation coefficient $\sim -0.09$). It is noteworthy that the regions containing the shocked molecular clouds do not always exhibit the lower $kT_\mathrm{e}$ values. These results, therefore, disfavor the pure thermal conduction scenario as the formation mechanism of the recombining plasma in G346.6$-$0.2.

We also would like to emphasize that the supernova explosion inside the stellar wind-bubble can naturally explain the adiabatic cooling scenario as the origin of recombining plasma in G346.6$-$0.2. Before the supernova explosion, a high-mass progenitor loses its gaseous envelope via stellar winds. The stellar winds (or mass loss events) can form not only a wind-blown bubble but also a dense CSM in the vicinity of the progenitor (e.g., \citeauthor{2013A&A...559A..69G} \citeyear{2013A&A...559A..69G} and references therein), and then the supernova explosion occurs inside the dense CSM. According to \cite{2012ApJ...750L..13M}, the dense CSM around red supergiants and Type IIn supernova progenitors can establish the CIE state of the plasma soon after the supernova explosion, and its remnant can evolve to the recombining plasma SNR via adiabatic cooling when the supernova shock breaks out of the CSM. In the case of G346.6$-$0.2, the presence of the stellar wind bubble is consistent with the scenario. If we take the relationship between the stellar mass and the wind-blown bubble size estimated by \citet{2013ApJ...769L..16C}, the progenitor mass of G346.6$-$0.2 is estimated to be $16~\mathrm{M_\odot}$ for the wind-blown bubble size of 10.6~pc. The progenitor mass is consistent with a massive red supergiant progenitor \citep{2015PASA...32...16S} which is suggested to be a progenitor of recombining plasma SNRs \citep{2012ApJ...750L..13M}. Further sensitive observations with high-spatial resolutions using ALMA will possibly unveil the remnant of the dense CSM toward the center of the SNR.

\begin{figure*}[]
\begin{center}
\includegraphics[width=140mm,clip]{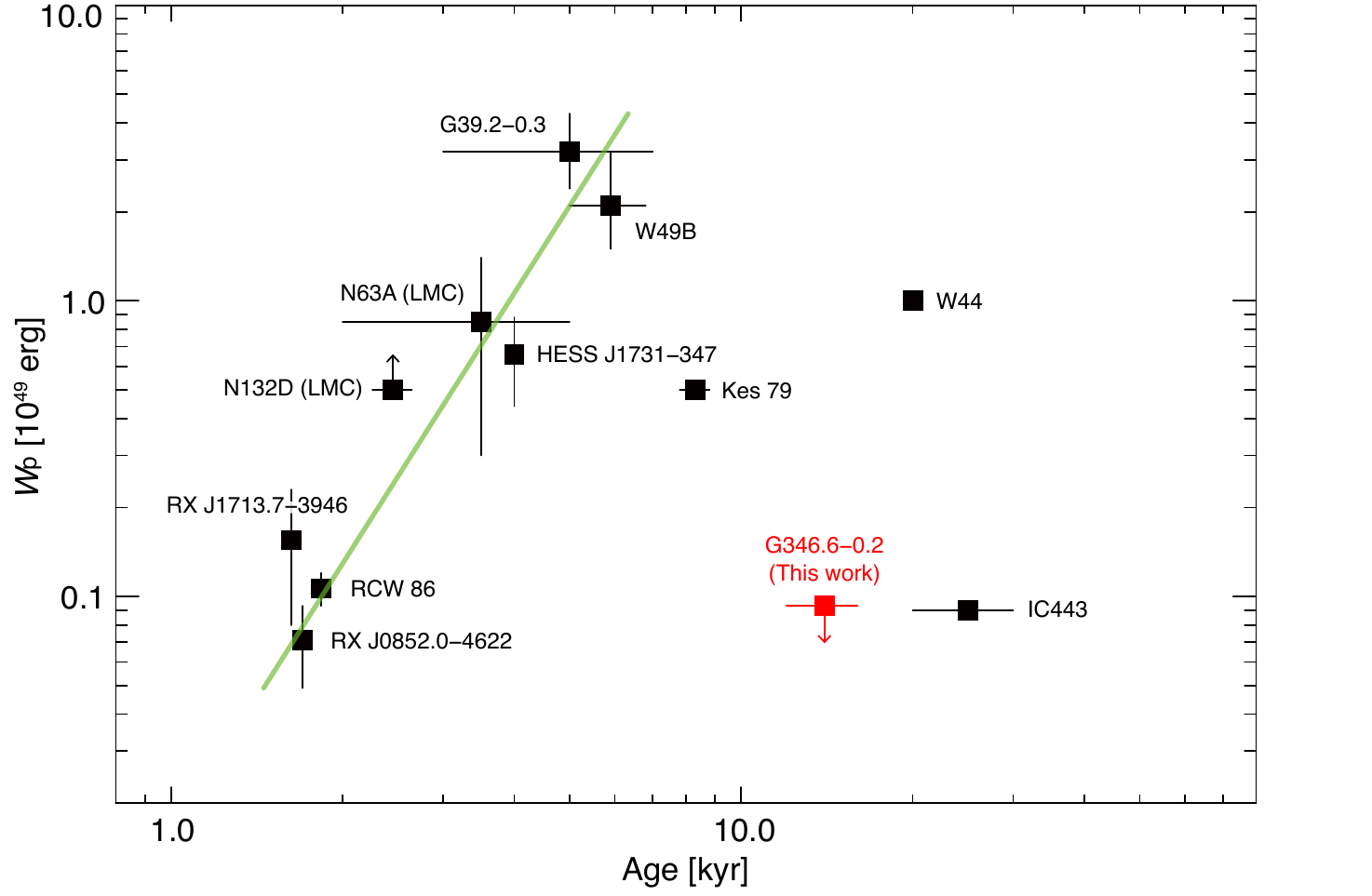}
\caption{Correlation plot between the age of SNRs and the total energy of cosmic-rays $W_\mathrm{p}$ \citep{2021arXiv210612009S}. The green line indicates the linear regression of the double-logarithmic plot applying the least-squares fitting for data points with the ages of SNRs below 6~kyr.}
\label{fig7}
\end{center}
\end{figure*}%

\subsection{Total Energy of Accelerated Cosmic-Rays}\label{subsec:cr_energy}
SNRs are believed to be the primary sources of Galactic cosmic rays, mainly consisting of protons, up to at least the energy of $\sim3$ PeV via the diffusive shock acceleration \citep[DSA, e.g.,][]{1978MNRAS.182..147B,1978ApJ...221L..29B}. Although the conventional value of total energy of accelerated cosmic rays $W_\mathrm{p}$ is thought to be $\sim10^{49}$--$10^{50}$ erg per a single supernova explosion, the observational constraint for $W_\mathrm{p}$ was insufficient because of lack of unified identification and/or quantification for shock interacting clouds using the CO/\ion{H}{1} radio line emission. Most recently, \cite{2021arXiv210612009S} presented the first reliable results of an SNR age--$W_\mathrm{p}$ relation for eleven gamma-ray SNRs. The authors discovered a positive correlation between the SNR age and $W_\mathrm{p}$ with ages below $\sim6$ kyr, suggesting that in-situ values of $W_\mathrm{p}$ for young SNRs are limited by the short duration time of cosmic-ray acceleration, also known as the age-limited acceleration \citep[cf.][]{2010A&A...513A..17O}. The older SNRs with ages more than $\sim8$ kyr, on the other hand, show a steady decrease of $W_\mathrm{p}$. This can be explained as an effect of the energy-dependent diffusion of cosmic rays \citep[e.g.,][]{1996A&A...309..917A,2007Ap&SS.309..365G}. In this section, we argue that G346.6$-$0.2 can be also naturally understood by the paradigm of the SNR age--$W_\mathrm{p}$ relation.

To derive the $W_\mathrm{p}$ value of G346.6$-$0.2, we first estimate the masses and number densities for the shock-interacting molecular and atomic clouds. The mass of molecular clouds $M_\mathrm{CO}$ can be estimated using the following equations:
\begin{eqnarray}
M_\mathrm{CO} = m_{\mathrm{p}} \mu \Omega D^2 \sum_{i} N_i(\mathrm{H}_2),\\
N(\mathrm{H}_2) = X \cdot W(\mathrm{CO}),
\label{eq1}
\end{eqnarray}
where $m_\mathrm{p}$ is the mass of atomic hydrogen, $\mu = 2.8$ is the mean molecular weight, $\Omega$ is the solid angle for each pixel, $D$ is the distance to the SNR, $N(\mathrm{H}_2)$ is the molecular hydrogen column density in units of cm$^{-2}$, $X$ is the CO-to-H$_2$ conversion factor of $2 \times 10^{20}$~cm$^{-2}$ (K~km~s$^{-1})^{-1}$ \citep{1993ApJ...416..587B}, and $W$(CO) is the velocity integrated intensity of $^{12}$CO($J$~=~1--0) emission line in units of K~km~s$^{-1}$. We estimated the mass of molecular clouds within the radio shell extent ($=$ shell radius $+$ $1/2$ shell thickness) to be $\sim4.1 \times 10^4$ $M_{\sun}$. The mass of atomic hydrogen clouds $M_\mathrm{HI}$ can be also given by the equations (4) and (5) under the optically thin assumption \citep{1990ARA&A..28..215D}: 
\begin{eqnarray}
M_\mathrm{HI} = m_{\mathrm{p}} \Omega D^2 \sum_{i} N_i(\mathrm{H{\textsc{i}}}),\\
N(\mathrm{H{\textsc{i}}}) = 1.823 \times W(\mathrm{H{\textsc{i}}}),
\label{eq2}
\end{eqnarray}
where $N$(\ion{H}{1}) is the atomic hydrogen column density in units of cm$^{-2}$ and $W$(\ion{H}{1}) is the velocity integrated intensity of \ion{H}{1} line emission in units of K~km~s$^{-1}$. We then estimated the mass of atomic clouds to be $\sim0.5 \times 10^4$ $M_{\sun}$, which is eight times less than the mass of molecular clouds. This trend is roughly consistent with other middle-aged SNRs \citep[e.g.,][]{2013ApJ...768..179Y,2021...submitted,2018ApJ...864..161K}. The averaged number densities within the wind-shell could be estimated to $\sim120$ cm$^{-3}$ for the molecular hydrogen $n$(H$_2$), and $\sim40$ cm$^{-3}$ for the atomic hydrogen $n$(\ion{H}{1}) by adopting the shell radius of 3\farcm29 ($\sim10.6$ pc) and the shell thickness of 1\farcm41 ($\sim4.6$ pc). Then we derived the number density of the total interstellar protons $n$ to be 2 $\times$ $n$(H$_2$) $+$ $n$(\ion{H}{1}) = 280 cm$^{-3}$.

The total energy of accelerated cosmic-rays $W_\mathrm{p}$ can be derived by using the following equation if the hadronic process is dominantly working \citep[e.g.,][]{2006A&A...449..223A}:
\begin{eqnarray}
W_\mathrm{p} \sim t_\mathrm{pp \rightarrow \pi^0} \times L_\gamma 
\label{eq3}
\end{eqnarray}
where $t_\mathrm{pp \rightarrow \pi^0} \sim 4.5 \times 10^{15}$ ($n$/1 cm$^{-3}$)$^{-1}$ s is the characteristic cooling time of cosmic-ray proton and $L_\gamma$ is the gamma-ray luminosity in units of erg s$^{-1}$. By adopting $L_\gamma$(1--100 GeV$) < 5.8 \times 10^{34}$ erg s$^{-1}$ and $n = 280$ cm$^{-3}$ at the SNR distance of 11.1 kpc, we finally derived $W_\mathrm{p}$(10--1000 GeV$) < 9.3 \times 10^{47}$ erg, corresponding to less than $\sim0.1$\% of the typical released kinetic energy of $\sim10^{51}$ erg of a supernova explosion.

Figure \ref{fig7} shows the SNR age--$W_\mathrm{p}$ diagram of twelve gamma-ray SNRs including G346.6$-$0.2. We can clearly see that G346.6$-$0.2 shows roughly consistent values with the other SNRs such as IC~443 \citep{2021...submitted}. Such a low energy amount suggests that most of the cosmic-rays in G346.6$-$0.2 have been escaped from the SNR, we can find the small in-situ value of $W_\mathrm{p}$. Further gamma-ray observations will possibly detect the hadronic gamma-ray emission due to the escaped cosmic-rays in the vicinity of the SNR. 

\section{Conclusions}\label{sec:conclusions}
We summarize the primary conclusions as follows:

\begin{enumerate}

\item We revealed CO and \ion{H}{1} clouds associated with the mixed-morphology SNR G346.6$-$0.2 using the NANTEN2 $^{12}$CO($J$~=~2--1) and ATCA \& Parkes \ion{H}{1} datasets. The \ion{H}{1} clouds are distributed surrounding the radio continuum shell except for the southeastern direction, while the CO clouds are located only the south and northeast shell of the SNR. The south CO clouds show a good spatial correspondence with the positions of the shock-excited 1720 MHz OH masers at $V_\mathrm{LSR} \sim -70$ km s$^{-1}$. The northeast CO cloud shows a high-kinetic temperature of $\sim 60$~K, suggesting that the shock heating occurred.

\item The cavity-like structure in the position-velocity diagram of \ion{H}{1} (possibly CO as well) indicates a wind-blown bubble which was likely formed by strong stellar winds from the high-mass progenitor of the SNR. The systemic velocity and expansion velocity of the wind bubble were derived to $\sim -72$ km s$^{-1}$ and $\sim 10$ km s$^{-1}$, respectively. The \ion{H}{1} absorption feature due to the SNR is seen in the velocity range from $-95$ to $-2$ km s$^{-1}$, suggesting that G346.6$-$0.2 is located on the far side of the Galactic center from us and the appropriate distance is to be $11.1^{+0.5}_{-0.3}$ kpc. The plasma age of the SNR was also revised to be $14 \pm 2$ kyr.

\item We found that the X-ray recombining plasma shows a clear spatial anti-correlation with the shocked CO clouds. In addition, there is no specific correlation between the CO intensities and the electron temperature of the recombining plasma. These results favor the adiabatic cooling scenario as the formation mechanism of recombining plasma in G346.6$-$0.2, rather than the thermal conduction scenario. This is also consistent with the presence of the stellar wind bubble because the dense CSM will be expected in the vicinity of the high-mass progenitor.  

\item Using the latest {\it{Fermi}}-LAT datasets, we placed a conservative upper limit of gamma-ray luminosity from the SNR G346.6$-$0.2 to $\sim5.8 \times 10^{34}$ erg s$^{-1}$ at the distance of 11.1 kpc. This corresponds to the total energy of cosmic-rays $W_\mathrm{p} < 9.3 \times 10^{47}$ erg adopting the target interstellar gas density of 280 cm$^{-3}$ if the hadronic process is dominantly working. The SNR age--$W_\mathrm{p}$ relation indicates that most of the accelerated cosmic-rays have been escaped from the SNR shell. 

\end{enumerate}

\section*{}
The NANTEN project is based on a mutual agreement between Nagoya University and the Carnegie Institution of Washington (CIW). We greatly appreciate the hospitality of all the staff members of the Las Campanas Observatory of CIW. We are thankful to many Japanese public donors and companies who contributed to the realization of the project. NANTEN2 is an international collaboration of 10 universities, Nagoya University, Osaka Prefecture University, University of Cologne, University of Bonn, Seoul National University, University of Chile, University of New South Wales, Macquarie University, University of Sydney, and ETH Zurich. The Mopra telescope, Australia Telescope Compact Array (ATCA), and the Parkes radio telescope are parts of the Australia Telescope National Facility which is funded by the Australian Government for operation as a National Facility managed by CSIRO. We acknowledge the Gamilaroi, Gomeroi, and Wiradjuri people as the traditional owners of the Observatory sites. The University of New South Wales Digital Filter Bank used for the observations with the Mopra Telescope was provided with support from the Australian Research Council. The Molonglo Observatory Synthesis Telescope (MOST) is operated by the University of Sydney with support from the Australian Research Council and the Science Foundation for Physics within the University of Sydney. This paper made use of information from the SEDIGISM survey database located at \href{https://sedigism.mpifr-bonn.mpg.de/index.html}{https://sedigism.mpifr-bonn.mpg.de/index.html}, which was constructed by James Urquhart and hosted by the Max Planck Institute for Radio Astronomy. Based on observations obtained with XMM-Newton, an ESA science mission with instruments and contributions directly funded by ESA Member States and NASA. This work was supported by JSPS KAKENHI Grant Numbers \href{https://kaken.nii.ac.jp/en/grant/KAKENHI-PUBLICLY-19H05075/}{JP19H05075} (H. Sano), \href{https://kaken.nii.ac.jp/en/grant/KAKENHI-PROJECT-20K14491/}{JP20K14491} (K. Nobukawa), \href{https://kaken.nii.ac.jp/en/grant/KAKENHI-PROJECT-20KK0071/}{JP20KK0071} (K. Nobukawa), \href{https://kaken.nii.ac.jp/en/grant/KAKENHI-PROJECT-21H01136/}{JP21H01136} (H. Sano), \href{https://kaken.nii.ac.jp/en/grant/KAKENHI-PROJECT-21K03615/}{JP21K03615} (K. Nobukawa), and \href{https://kaken.nii.ac.jp/en/grant/KAKENHI-PROJECT-21J00031/}{JP21J00031} (H. Suzuki). K. Nobukawa was also supported by Yamada Science Foundation.

\software{IDL Astronomy User's Library \citep{1993ASPC...52..246L}, MIRIAD \citep{1995ASPC...77..433S}, SAS \citep[v19.1.0:][]{2004ASPC..314..759G}, ESAS \citep[][]{2008A&A...478..575K}, HEAsoft \citep[v6.28:][]{2014ascl.soft08004N}, Fermitools (v1.2.23; \url{https://github.com/fermi-lat/Fermitools-conda/})}

\facilities{NANTEN2, Mopra, Atacama Pathfinder Experiment (APEX), Australia Telescope Compact Array (ATCA), Parkes, Molonglo Observatory Synthesis Telescope (MOST), {\it{XMM-Newton}}, and {\it Fermi}-LAT}


\begin{thebibliography}{99}
\bibitem[Abdo et al.(2009)]{2009ApJ...706L...1A} Abdo, A.~A., Ackermann, M., Ajello, M., et al.\ 2009, \apjl, 706, L1. doi:10.1088/0004-637X/706/1/L1
\bibitem[Aharonian \& Atoyan(1996)]{1996A&A...309..917A} Aharonian, F.~A. \& Atoyan, A.~M.\ 1996, \aap, 309, 917
\bibitem[Aharonian et al.(2006)]{2006A&A...449..223A} Aharonian, F., Akhperjanian, A.~G., Bazer-Bachi, A.~R., et al.\ 2006, \aap, 449, 223. doi:10.1051/0004-6361:20054279
\bibitem[Anderl et al.(2014)]{2014A&A...569A..81A} Anderl, S., Gusdorf, A., \& G{\"u}sten, R.\ 2014, \aap, 569, A81. doi:10.1051/0004-6361/201423561
\bibitem[Andersen et al.(2011)]{2011ApJ...742....7A} Andersen, M., Rho, J., Reach, W.~T., et al.\ 2011, \apj, 742, 7. doi:10.1088/0004-637X/742/1/7
\bibitem[Arikawa et al.(1999)]{1999PASJ...51L...7A} Arikawa, Y., Tatematsu, K., Sekimoto, Y., et al.\ 1999, \pasj, 51, L7. doi:10.1093/pasj/51.4.L7
\bibitem[Auchettl et al.(2017)]{2017ApJ...847..121A} Auchettl, K., Ng, C.-Y., Wong, B.~T.~T., et al.\ 2017, \apj, 847, 121. doi:10.3847/1538-4357/aa830e
\bibitem[Bell(1978)]{1978MNRAS.182..147B} Bell, A.~R.\ 1978, \mnras, 182, 147. doi:10.1093/mnras/182.2.147
\bibitem[Bertsch et al.(1993)]{1993ApJ...416..587B} Bertsch, D.~L., Dame, T.~M., Fichtel, C.~E., et al.\ 1993, \apj, 416, 587. doi:10.1086/173261
\bibitem[Blake et al.(1987)]{1987ApJ...315..621B} Blake, G.~A., Sutton, E.~C., Masson, C.~R., et al.\ 1987, \apj, 315, 621. doi:10.1086/165165
\bibitem[Blandford \& Ostriker(1978)]{1978ApJ...221L..29B} Blandford, R.~D. \& Ostriker, J.~P.\ 1978, \apjl, 221, L29. doi:10.1086/182658
\bibitem[Braiding et al.(2018)]{2018PASA...35...29B} Braiding, C., Wong, G.~F., Maxted, N.~I., et al.\ 2018, \pasa, 35, e029. doi:10.1017/pasa.2018.18
\bibitem[Brand \& Blitz(1993)]{1993A&A...275...67B} Brand, J. \& Blitz, L.\ 1993, \aap, 275, 67
\bibitem[Castro \& Slane(2010)]{2010ApJ...717..372C} Castro, D. \& Slane, P.\ 2010, \apj, 717, 372. doi:10.1088/0004-637X/717/1/372
\bibitem[Denoyer(1979)]{1979ApJ...232L.165D} Denoyer, L.~K.\ 1979, \apjl, 232, L165. doi:10.1086/183057
\bibitem[Chen et al.(2013)]{2013ApJ...769L..16C} Chen, Y., Zhou, P., \& Chu, Y.-H.\ 2013, \apjl, 769, L16. doi:10.1088/2041-8205/769/1/L16
\bibitem[Dickey \& Lockman(1990)]{1990ARA&A..28..215D} Dickey, J.~M. \& Lockman, F.~J.\ 1990, \araa, 28, 215. doi:10.1146/annurev.aa.28.090190.001243
\bibitem[Dubner et al.(1993)]{1993AJ....105.2251D} Dubner, G.~M., Moffett, D.~A., Goss, W.~M., et al.\ 1993, \aj, 105, 2251. doi:10.1086/116603
\bibitem[Elitzur(1976)]{1976ApJ...203..124E} Elitzur, M.\ 1976, \apj, 203, 124. doi:10.1086/154054
\bibitem[Ergin \& Ercan(2012)]{2012AIPC.1505..265E} Ergin, T. \& Ercan, E.~N.\ 2012, High Energy Gamma-Ray Astronomy: 5th International Meeting on High Energy Gamma-Ray Astronomy, 1505, 265. doi:10.1063/1.4772248
\bibitem[Fukui(2008)]{2008AIPC.1085..104F} Fukui, Y.\ 2008, American Institute of Physics Conference Series, 1085, 104. doi:10.1063/1.3076625
\bibitem[Fukui et al.(2012)]{2012ApJ...746...82F} Fukui, Y., Sano, H., Sato, J., et al.\ 2012, \apj, 746, 82. doi:10.1088/0004-637X/746/1/82
\bibitem[Gabici et al.(2007)]{2007Ap&SS.309..365G} Gabici, S., Aharonian, F.~A., \& Blasi, P.\ 2007, \apss, 309, 365. doi:10.1007/s10509-007-9427-6
\bibitem[Gabriel et al.(2004)]{2004ASPC..314..759G} Gabriel, C., Denby, M., Fyfe, D.~J., et al.\ 2004, Astronomical Data Analysis Software and Systems (ADASS) XIII, 314, 759
\bibitem[Georgy et al.(2013)]{2013A&A...559A..69G} Georgy, C., Walder, R., Folini, D., et al.\ 2013, \aap, 559, A69. doi:10.1051/0004-6361/201321226
\bibitem[Goldreich \& Kwan(1974)]{1974ApJ...189..441G} Goldreich, P. \& Kwan, J.\ 1974, \apj, 189, 441. doi:10.1086/152821
\bibitem[Hachisu et al.(1996)]{1996ApJ...470L..97H} Hachisu, I., Kato, M., \& Nomoto, K.\ 1996, \apjl, 470, L97. doi:10.1086/310303
\bibitem[Hachisu et al.(1999a)]{1999ApJ...519..314H} Hachisu, I., Kato, M., Nomoto, K., et al.\ 1999a, \apj, 519, 314. doi:10.1086/307370
\bibitem[Hachisu et al.(1999b)]{1999ApJ...522..487H} Hachisu, I., Kato, M., \& Nomoto, K.\ 1999b, \apj, 522, 487. doi:10.1086/307608
\bibitem[Itoh \& Masai(1989)]{1989MNRAS.236..885I} Itoh, H. \& Masai, K.\ 1989, \mnras, 236, 885. doi:10.1093/mnras/236.4.885
\bibitem[Kawasaki et al.(2002)]{2002ApJ...572..897K} Kawasaki, M.~T., Ozaki, M., Nagase, F., et al.\ 2002, \apj, 572, 897. doi:10.1086/340383
\bibitem[Kawasaki et al.(2005)]{2005ApJ...631..935K} Kawasaki, M., Ozaki, M., Nagase, F., et al.\ 2005, \apj, 631, 935. doi:10.1086/432591
\bibitem[Kerr \& Lynden-Bell(1986)]{1986MNRAS.221.1023K} Kerr, F.~J. \& Lynden-Bell, D.\ 1986, \mnras, 221, 1023. doi:10.1093/mnras/221.4.1023
\bibitem[Koo et al.(1990)]{1990ApJ...364..178K} Koo, B.-C., Reach, W.~T., Heiles, C., et al.\ 1990, \apj, 364, 178. doi:10.1086/169400
\bibitem[Koo \& Heiles(1991)]{1991ApJ...382..204K} Koo, B.-C. \& Heiles, C.\ 1991, \apj, 382, 204. doi:10.1086/170709
\bibitem[Koralesky et al.(1998)]{1998AJ....116.1323K} Koralesky, B., Frail, D.~A., Goss, W.~M., et al.\ 1998, \aj, 116, 1323. doi:10.1086/300508
\bibitem[Kuntz \& Snowden(2008)]{2008A&A...478..575K} Kuntz, K.~D. \& Snowden, S.~L.\ 2008, \aap, 478, 575. doi:10.1051/0004-6361:20077912
\bibitem[Kuriki et al.(2018)]{2018ApJ...864..161K} Kuriki, M., Sano, H., Kuno, N., et al.\ 2018, \apj, 864, 161. doi:10.3847/1538-4357/aad7be
\bibitem[Landecker et al.(1989)]{1989MNRAS.237..277L} Landecker, T.~L., Pineault, S., Routledge, D., et al.\ 1989, \mnras, 237, 277. doi:10.1093/mnras/237.1.277
\bibitem[Landsman(1993)]{1993ASPC...52..246L} Landsman, W.~B.\ 1993, Astronomical Data Analysis Software and Systems II, 52, 246
\bibitem[Langer \& Penzias(1990)]{1990ApJ...357..477L} Langer, W.~D. \& Penzias, A.~A.\ 1990, \apj, 357, 477. doi:10.1086/168935
\bibitem[Leahy \& Tian(2010)]{2010ASPC..438..365L} Leahy, D. \& Tian, W.\ 2010, The Dynamic Interstellar Medium: A Celebration of the Canadian Galactic Plane Survey, 438, 365
\bibitem[McClure-Griffiths et al.(2005)]{2005ApJS..158..178M} McClure-Griffiths, N.~M., Dickey, J.~M., Gaensler, B.~M., et al.\ 2005, \apjs, 158, 178. doi:10.1086/430114
\bibitem[Masai(1994)]{1994ApJ...437..770M} Masai, K.\ 1994, \apj, 437, 770. doi:10.1086/175037
\bibitem[Matsumura et al.(2017)]{2017ApJ...851...73M} Matsumura, H., Tanaka, T., Uchida, H., et al.\ 2017, \apj, 851, 73. doi:10.3847/1538-4357/aa9bdf
\bibitem[Maxted et al.(2012)]{2012MNRAS.422.2230M} Maxted, N.~I., Rowell, G.~P., Dawson, B.~R., et al.\ 2012, \mnras, 422, 2230. doi:10.1111/j.1365-2966.2012.20766.x
\bibitem[Maxted et al.(2013)]{2013PASA...30...55M} Maxted, N.~I., Rowell, G.~P., Dawson, B.~R., et al.\ 2013, \pasa, 30, e055. doi:10.1017/pasa.2013.35
\bibitem[Moriya(2012)]{2012ApJ...750L..13M} Moriya, T.~J.\ 2012, \apjl, 750, L13. doi:10.1088/2041-8205/750/1/L13
\bibitem[Nasa High Energy Astrophysics Science Archive Research Center(2014)]{2014ascl.soft08004N} Nasa High Energy Astrophysics Science Archive Research Center (Heasarc)\ 2014, Astrophysics Source Code Library. ascl:1408.004
\bibitem[Ohira et al.(2010)]{2010A&A...513A..17O} Ohira, Y., Murase, K., \& Yamazaki, R.\ 2010, \aap, 513, A17. doi:10.1051/0004-6361/200913495
\bibitem[Okon et al.(2018)]{2018PASJ...70...35O} Okon, H., Uchida, H., Tanaka, T., et al.\ 2018, \pasj, 70, 35. doi:10.1093/pasj/psy022
\bibitem[Okon et al.(2020)]{2020ApJ...890...62O} Okon, H., Tanaka, T., Uchida, H., et al.\ 2020, \apj, 890, 62. doi:10.3847/1538-4357/ab6987
\bibitem[Pavlovic et al.(2014)]{2014SerAJ.189...25P} Pavlovic, M.~Z., Dobardzic, A., Vukotic, B., et al.\ 2014, Serbian Astronomical Journal, 189, 25. doi:10.2298/SAJ1489025P
\bibitem[Ranasinghe \& Leahy(2017)]{2017ApJ...843..119R} Ranasinghe, S. \& Leahy, D.~A.\ 2017, \apj, 843, 119. doi:10.3847/1538-4357/aa7894
\bibitem[Reach et al.(2006)]{2006AJ....131.1479R} Reach, W.~T., Rho, J., Tappe, A., et al.\ 2006, \aj, 131, 1479. doi:10.1086/499306
\bibitem[Rho \& Petre(1998)]{1998ApJ...503L.167R} Rho, J. \& Petre, R.\ 1998, \apjl, 503, L167. doi:10.1086/311538
\bibitem[Sano et al.(2010)]{2010ApJ...724...59S} Sano, H., Sato, J., Horachi, H., et al.\ 2010, \apj, 724, 59. doi:10.1088/0004-637X/724/1/59
\bibitem[Sano et al.(2013)]{2013ApJ...778...59S} Sano, H., Tanaka, T., Torii, K., et al.\ 2013, \apj, 778, 59. doi:10.1088/0004-637X/778/1/59
\bibitem[Sano et al.(2017)]{2017JHEAp..15....1S} Sano, H., Reynoso, E.~M., Mitsuishi, I., et al.\ 2017, Journal of High Energy Astrophysics, 15, 1. doi:10.1016/j.jheap.2017.04.002
\bibitem[Sano et al.(2018)]{2018ApJ...867....7S} Sano, H., Yamane, Y., Tokuda, K., et al.\ 2018, \apj, 867, 7. doi:10.3847/1538-4357/aae07c
\bibitem[Sano et al.(2019)]{2019ApJ...881...85S} Sano, H., Matsumura, H., Yamane, Y., et al.\ 2019, \apj, 881, 85. doi:10.3847/1538-4357/ab2ade
\bibitem[Sano et al.(2021)]{2021arXiv210612009S} Sano, H., Yoshiike, S., Yamane, Y., et al.\ 2021, arXiv:2106.12009
\bibitem[Sault et al.(1995)]{1995ASPC...77..433S} Sault, R.~J., Teuben, P.~J., \& Wright, M.~C.~H.\ 1995, Astronomical Data Analysis Software and Systems IV, 77, 433
\bibitem[Schneider et al.(1998)]{1998A&A...335.1049S} Schneider, N., Stutzki, J., Winnewisser, G., et al.\ 1998, \aap, 335, 1049
\bibitem[Schuller et al.(2021)]{2021MNRAS.500.3064S} Schuller, F., Urquhart, J.~S., Csengeri, T., et al.\ 2021, \mnras, 500, 3064. doi:10.1093/mnras/staa2369
\bibitem[Scoville \& Solomon(1974)]{1974ApJ...187L..67S} Scoville, N.~Z. \& Solomon, P.~M.\ 1974, \apjl, 187, L67. doi:10.1086/181398
\bibitem[Seta et al.(1998)]{1998ApJ...505..286S} Seta, M., Hasegawa, T., Dame, T.~M., et al.\ 1998, \apj, 505, 286. doi:10.1086/306141
\bibitem[Sezer et al.(2011)]{2011MNRAS.415..301S} Sezer, A., G{\"o}k, F., Hudaverdi, M., et al.\ 2011, \mnras, 415, 301. doi:10.1111/j.1365-2966.2011.18710.x
\bibitem[Smartt(2015)]{2015PASA...32...16S} Smartt, S.~J.\ 2015, \pasa, 32, e016. doi:10.1017/pasa.2015.17
\bibitem[Suzuki et al.(2021)]{2021ApJ...914..103S} Suzuki, H., Bamba, A., \& Shibata, S.\ 2021, \apj, 914, 103. doi:10.3847/1538-4357/abfb02
\bibitem[Tanaka et al.(2011)]{2011ApJ...740L..51T} Tanaka, T., Allafort, A., Ballet, J., et al.\ 2011, \apjl, 740, L51. doi:10.1088/2041-8205/740/2/L51
\bibitem[Weaver et al.(1977)]{1977ApJ...218..377W} Weaver, R., McCray, R., Castor, J., et al.\ 1977, \apj, 218, 377. doi:10.1086/155692
\bibitem[Whiteoak \& Green(1996)]{1996A&AS..118..329W} Whiteoak, J.~B.~Z. \& Green, A.~J.\ 1996, \aaps, 118, 329
\bibitem[Wootten(1977)]{1977ApJ...216..440W} Wootten, H.~A.\ 1977, \apj, 216, 440. doi:10.1086/155485
\bibitem[Wootten(1981)]{1981ApJ...245..105W} Wootten, A.\ 1981, \apj, 245, 105. doi:10.1086/158790
\bibitem[Yamaguchi(2020)]{2020AN....341..150Y} Yamaguchi, H.\ 2020, Astronomische Nachrichten, 341, 150. doi:10.1002/asna.202023771
\bibitem[Yamauchi et al.(2008)]{2008PASJ...60.1143Y} Yamauchi, S., Ueno, M., Koyama, K., et al.\ 2008, \pasj, 60, 1143. doi:10.1093/pasj/60.5.1143
\bibitem[Yamauchi et al.(2013)]{2013PASJ...65....6Y} Yamauchi, S., Nobukawa, M., Koyama, K., et al.\ 2013, \pasj, 65, 6. doi:10.1093/pasj/65.1.6
\bibitem[Yoshiike et al.(2013)]{2013ApJ...768..179Y} Yoshiike, S., Fukuda, T., Sano, H., et al.\ 2013, \apj, 768, 179. doi:10.1088/0004-637X/768/2/179
\bibitem[Yoshiike et al.(2021)]{2021...submitted} Yoshiike, S., Sano, H., Fukuda, T.,  et al.\ 2021, to be submitted
\bibitem[Zhou et al.(2016)]{2016ApJ...826...34Z} Zhou, P., Chen, Y., Zhang, Z.-Y., et al.\ 2016, \apj, 826, 34. doi:10.3847/0004-637X/826/1/34
\end{thebibliography}
\end{document}